\documentclass[twocolumn]{jpsj3}
\usepackage{graphicx}
\usepackage{wrapfig}
\usepackage{lastpage}

\usepackage{subfigure}

\title{Fragmentation of a viscoelastic food by human mastication}

\author{Naoki \textsc{Kobayashi}\thanks{E-mail address: knaoki@phys.chuo-u.ac.jp}, Kaoru \textsc{Kohyama}$^1$, Kouichi \textsc{Shiozawa}$^2$}

\inst{
 Department of Physics, Chuo University, Kasuga, Bunkyo-ku, Tokyo 112-8551\\
$^1$ Food Physics Laboratory, National Food Research Institute, National Agriculture and Food Research
Organization, Kannondai, Tsukuba, Ibaraki 305-8642\\
$^2$ Department of Physiology, Tsurumi University School of Dental Medicine, Tsurumi,Tsurumi-ku, Yokohama-shi, Kanagawa 230-8501
}

\recdate{}

\abst{
Fragment-size distributions have been studied experimentally in masticated viscoelastic food (fish sausage).
The mastication experiment in seven subjects was examined.
We classified the obtained results into two groups, namely, a single lognormal distribution
 group and a lognormal distribution with exponential tail group.
The facts suggest that the individual variability might affect the fragmentation pattern when
 the food sample has a much more complicated physical property.
In particular, the latter result (lognormal distribution with exponential tail) indicates that
 the fragmentation pattern by human mastication for fish sausage is different from the fragmentation
 pattern for raw carrot shown in our previous study \cite{KKSM06}.
The excellent data fitting by the lognormal distribution with exponential tail implies
 that the fragmentation process has a size-segregation-structure between large and small
 parts.
In order to explain this structure, we propose a mastication model for fish sausage
 based on stochastic processes.
}

\kword{%
human mastication, viscoelastic food, lognormal distribution
}

\begin{document}

\sloppy
\maketitle

\section{Introduction}

Mastication is an in-mouth fragmentation process in which food is broken, 
 ground or crushed by the teeth to prepare for swallowing and digestion \cite{Bourne03}.
In general, a major problem of the oral process
 observation analysis is lack of direct visualization of the process itself.
This fact indicates that by investigating the food status
 before and after eating experimentally and numerically, we
 can understand some of the principal features of mastication
 processes and propose some phenomenological models \cite{LPAB02}.
 
As the first approximation, we propose that mastication is
 a sequential fragmentation in the oral cavity between the teeth and food. 
Fragmentation is a very complicated phenomenon, and it is difficult to
 understand its dynamics.
One of methods used to understand the fragmentation process is
 to investigate a fragment-size distribution.
The fragment-size distribution resulting from various types of fracture 
 has attracted the interest of physicists for many years.
We can give a few examples for the studies of the fragment-size distributions,
 the studies of astelloids \cite{Klacka92}, glass rods \cite{IM92}, glass
 plates \cite{KSH03}, egg-shells \cite{WKHK04} and so on.

In dental science, there has been a considerable study of the
 fragment-size distribution \cite{OBBK84,BAMH93,PMW04,JMMDWP07}.
We also experimentally studied the fragment-size distribution
 by masticating raw carrots, which are regarded as a brittle material \cite{KKSM06}.
We reported that a lognormal distribution well fits the entire region
 for masticated fragments of raw carrots for several chewing
 strokes.
The above result indicates that the fragmentation of raw carrots by human
 mastication is characterized by the
 effect of random multiplicative stochastic processes in statistics \cite{CS88}.

Solid foods which need mastication by the teeth to eat can be categorized into four
 groups by means of bite-force curve observed in the first chew of humans \cite{KSA01}.
They are A) sponge-like foods (such as bread and sponge cake), B) gels (such as agar
 jelly, cooked rice, and fish gel), C) wet crisp foods (such as raw carrot, apple and
 cucumber), and D) dry crisp foods (such as cracker and cookie).
As raw carrot belonging in the group C was studied in our previous paper \cite{KKSM06},
 we choose fish sausage from the gel group.
Generally, gels are viscoelastic unlike crisp foods.
The mechanical characteristics of viscoelastic gels were soft, easy to deform.
Therefore, the viscoelastic food are even chewable for the elderly \cite{SSM07}.
On the other hand, they are difficult to break off completely \cite{KHDS05, KNYYHS07},
 though it is homogeneous and isotropic nature.
We have a little information for the fragmentation process of viscoelastic food.
The gel structure of fish sausage formed by fish protein molecular networks.

Among the four groups discussed above, saliva absorption effect during mastication may
 be greater in sponge-like food which has porous structure ready to hold liquid, and
 in dry-crisp food with high water absorption capacity.
However since gel type food and wet-crisp foods are rich in water, no significant saliva
 absorption may occur.
In the model proposed by Hutchings and Lillford \cite{HL88}, mastication is the process
 to break down the food structure and to lubricate the bolus surface ready to swallow.
Evidently, the former is the main effect required for fish sausage and raw carrot.
To discuss the differences in fragmentation by human mastication between those two groups
 of food is valuable.

In this paper, we present an experimental study of the fragmentation of
 fish sausage by human mastication.
So we examine data fittings by various distributions.
Finally, we discuss a plausible process of fragmentation of fish sausage
 by mastication in accordance with physical viewpoints.

\section{Materials and Methods}
{\it Test food and subjects}.
We used fish sausage (Osakanano-sausage, Nippon Suisan Kaisha, Lid., JAPAN) as test
 food in this study. 
The test food was cut into a cylinder (diameter and height; $24$ mm and $15$ mm,
 respectively) of about $7$ g. 
Seven healthy subjects (4 males and 3 females, mean age 26.4 years) voluntarily
 participated in this study. 
They had natural dentition without severe malocclusion and periodontal disease.
Written informed consent was obtained from each subject after full explanation of
 the experiment.\\
\\
{\it Collect of food fragments}.
First, each subject masticated the test food until swallowing as usual to count
 the number of chewing cycles until swallowing.
Next, we calculated the individual number of chewing cycles until the halfway of
 the mastication.
To collect the food bolus at the halfway of the mastication and just before
 swallowing, subjects were asked to masticate the test food until individually
 prescribed number of chewing cycles, and they spat the bolus into a beaker.
The collection of food bolus was performed two times per subject and each condition.
In order to expectorate entirely the subject rinsed their mouth with water.
The food fragments and water were carefully stirred in a beaker with a grass rod
 and pass through a sieve with a mesh size of $0.5$ mm.
After fine fragments were washed through the sieve with running water, the fragments
 on the sieve were spread evenly on the transparent acrylic board ($300$ mm $\times$ $300$ mm).
Then, we made a copy of this board using copy machine (DocuCentra-II, FUJI XEROX, JAPAN)
 without cover and stored the copy on a personal computer at a suitable resolution
(about $0.13$ mm/pixel).\\
\\
{\it Statistics}.
The additional data from second trials was then assimilated and sorted
 in terms of size from the largest one, i.e., cumulative number.
Then, we obtained the cumulative size distribution for each number of 
 chewing strokes and applied a suitable distribution to fit a curve
 to the data in each case.
For fitting the curve, we performed a nonlinear least-square method using R
 (version 2.4.1 for windows) \cite{R}.

\section{Results and Discussion}

\begin{figure}[h]
\centering
\subfigure[]{
\includegraphics*[width=0.42\linewidth]{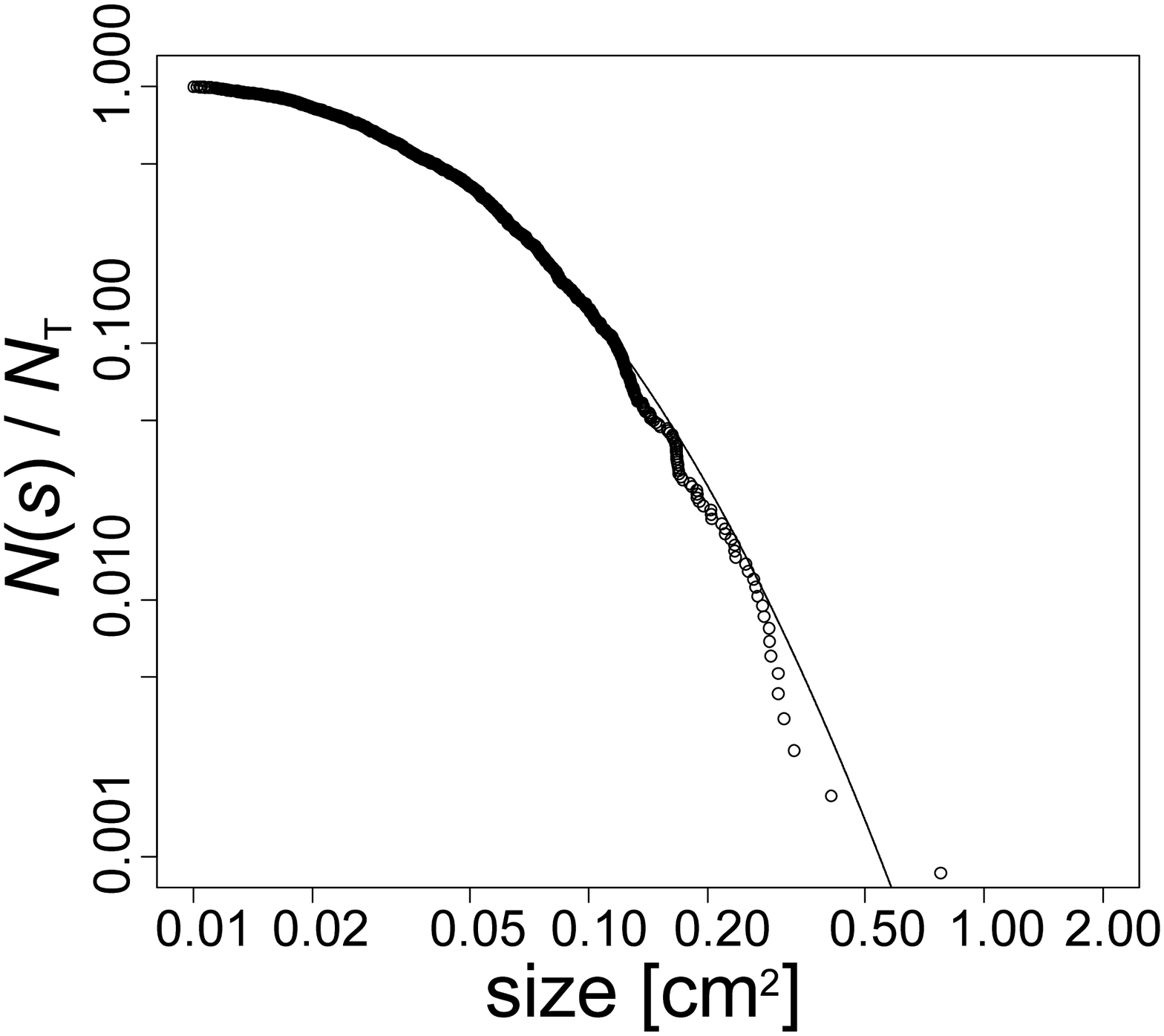}
\label{fig:Fig1a}}
\hfill
\subfigure[]{
\includegraphics*[width=0.42\linewidth]{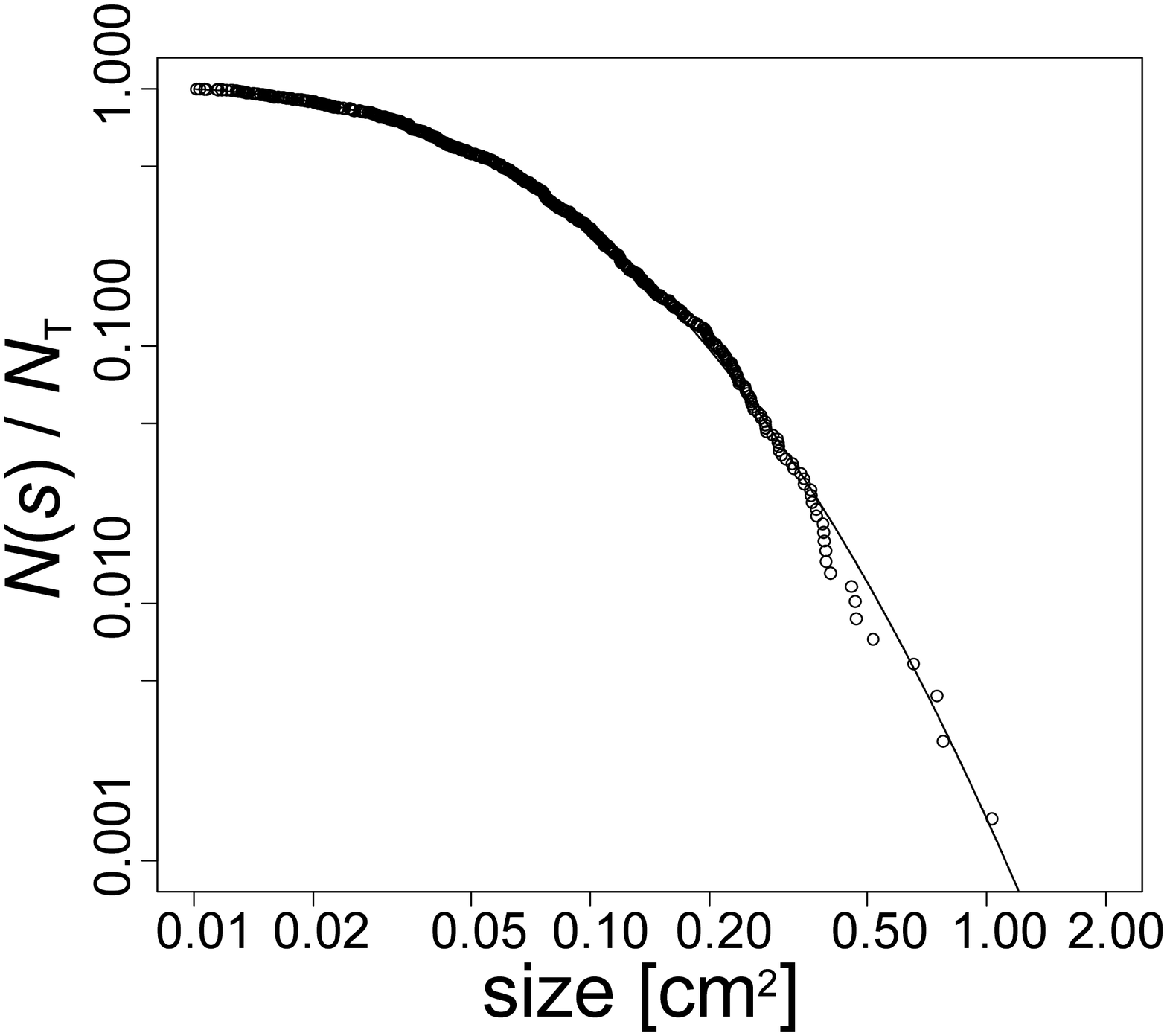}
\label{fig:Fig1b}}
\caption{
Log-log plots for the cumulative numberof masticated food
 fragments of fish sausage after (a) 32 and (b) 16 chews.
The solid line indecates a lognormal distribution for (a) $N_{\rm{T}} = 1161,
 \bar{s} = 0.0375, \sigma = 0.859$ and $N / N_{\rm{T}} = 1.08$, and
 (b) $N_{\rm{T}} = 688, \bar{s} = 0.0559, \sigma = 0.965$ and
 $N / N_{\rm{T}} = 1.04$.
}
\label{fig:Fig1}
\end{figure}
\begin{figure}[h]
\centering
\subfigure[]{
\includegraphics*[width=0.42\linewidth]{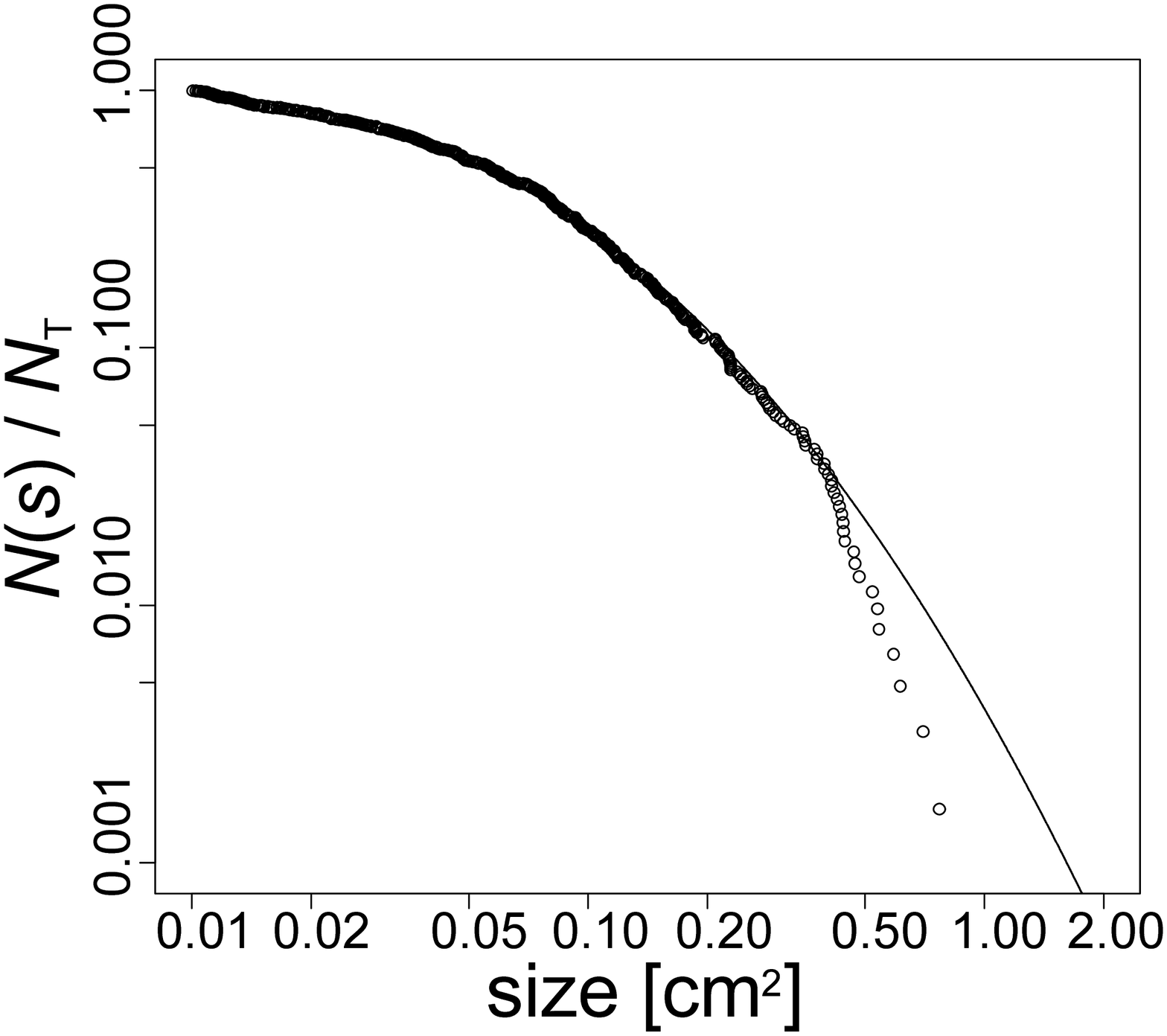}
\label{fig:Fig2a}}
\hfill
\subfigure[]{
\includegraphics*[width=0.42\linewidth]{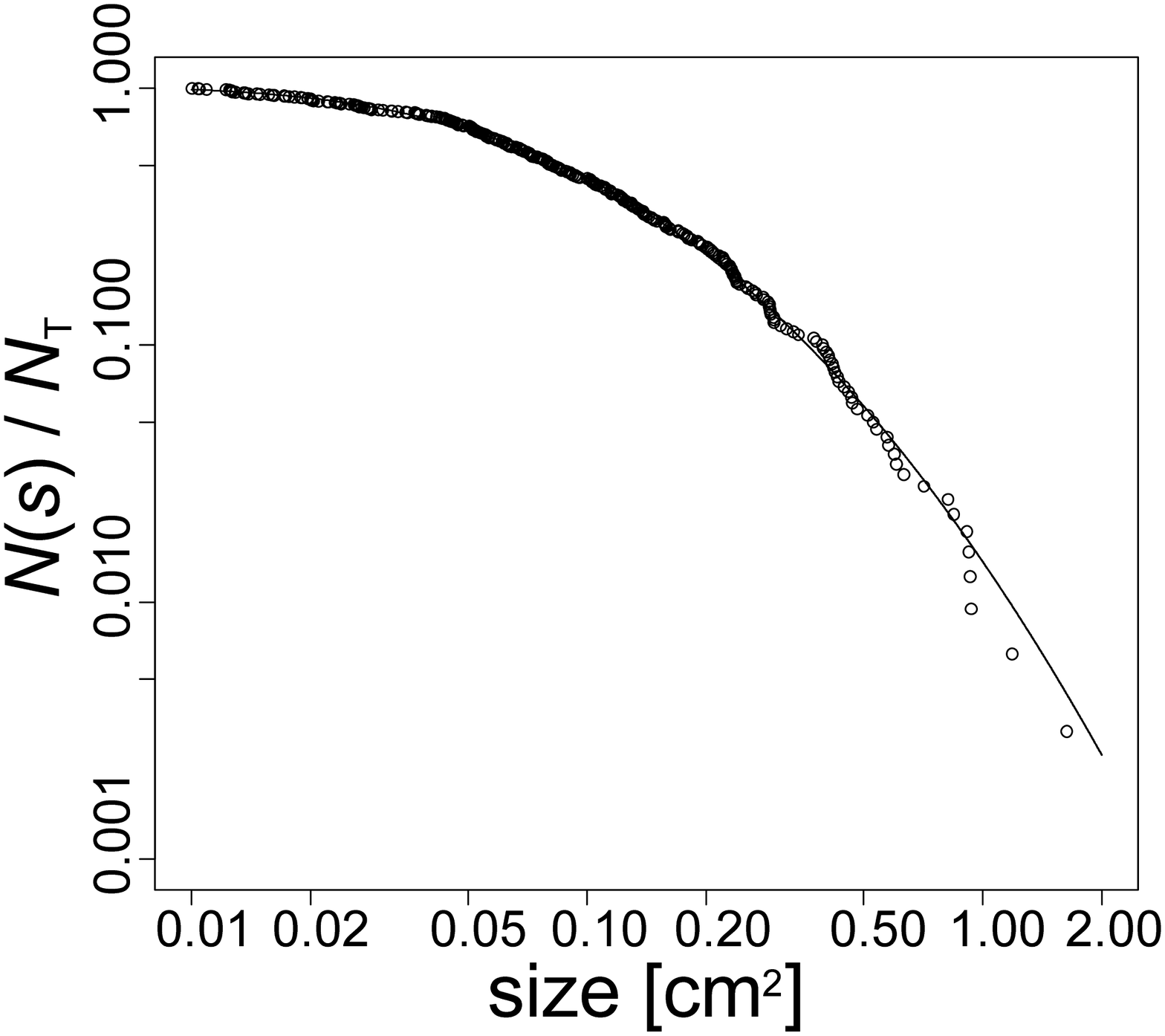}
\label{fig:Fig2b}}
\caption{
Same as Fig. \ref{fig:Fig1} except for the number of chewing
 strokes, i.e., (a) 24 and (b) 12.
The fitting parameters are (a) $N_{\rm{T}} = 620,
 \bar{s} = 0.0535, \sigma = 1.10$ and $N / N_{\rm{T}} = 1.02$, and
 (b) $N_{\rm{T}} = 319, \bar{s} = 0.0837, \sigma = 1.13$ and
 $N / N_{\rm{T}} = 1.02$.
}
\label{fig:Fig2}
\end{figure}
\begin{figure}[h]
\centering
\subfigure[]{
\includegraphics*[width=0.42\linewidth]{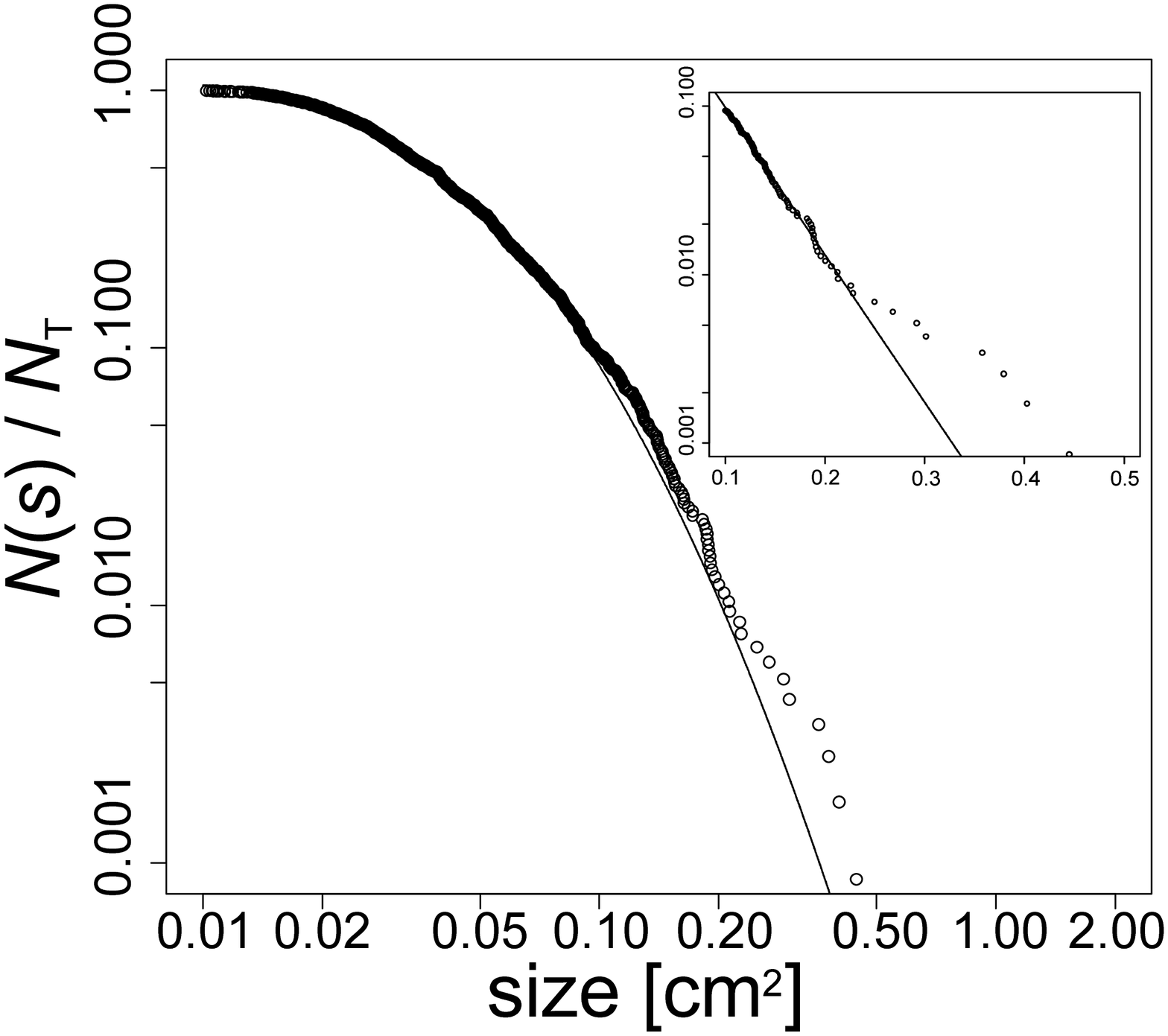}
\label{fig:Fig3a}}
\hfill
\subfigure[]{
\includegraphics*[width=0.42\linewidth]{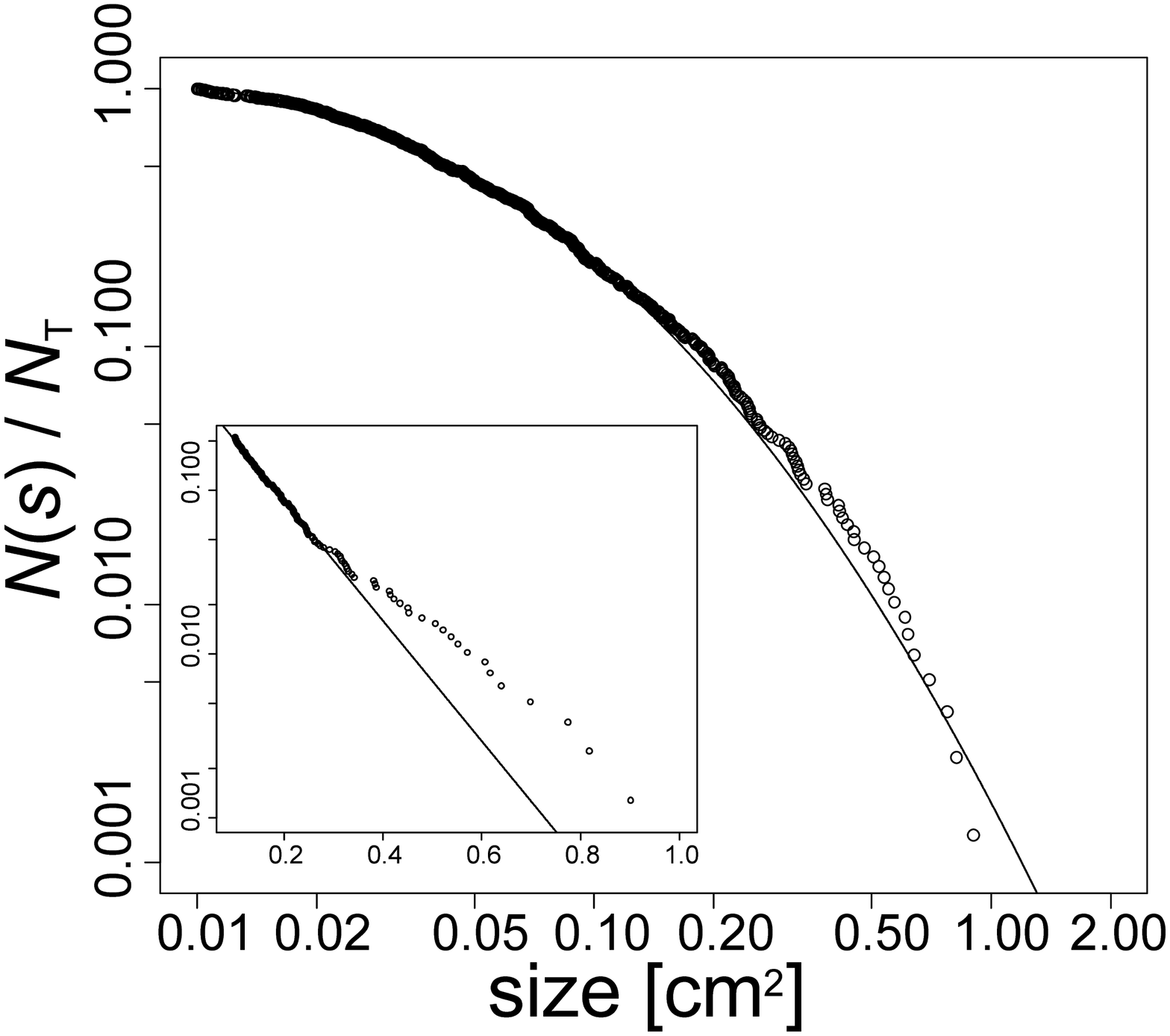}
\label{fig:Fig3b}}
\caption{
Same as Fig. \ref{fig:Fig1} except for the numner of chewing
 strokes, i.e., (a) 50 and (b) 25.
The fitting parameters are (a) $N_{\rm{T}} = 1162,
 \bar{s} = 0.0341, \sigma = 0.754$ and $N / N_{\rm{T}} = 1.11$, and (b)
 $N_{\rm{T}} = 784, \bar{s} = 0.0369, \sigma = 1.11$ and $N / N_{\rm{T}} = 1.15$.
The insets show semi-log plots for the tail part. 
The fitting parameters are (a) $A = 0.732$, $B = 0.0497$ and 
 (b) $A = 0.474$, $B = 0.118$.
}
\label{fig:Fig3}
\end{figure}
\begin{figure}[h]
\centering
\subfigure[]{
\includegraphics*[width=0.42\linewidth]{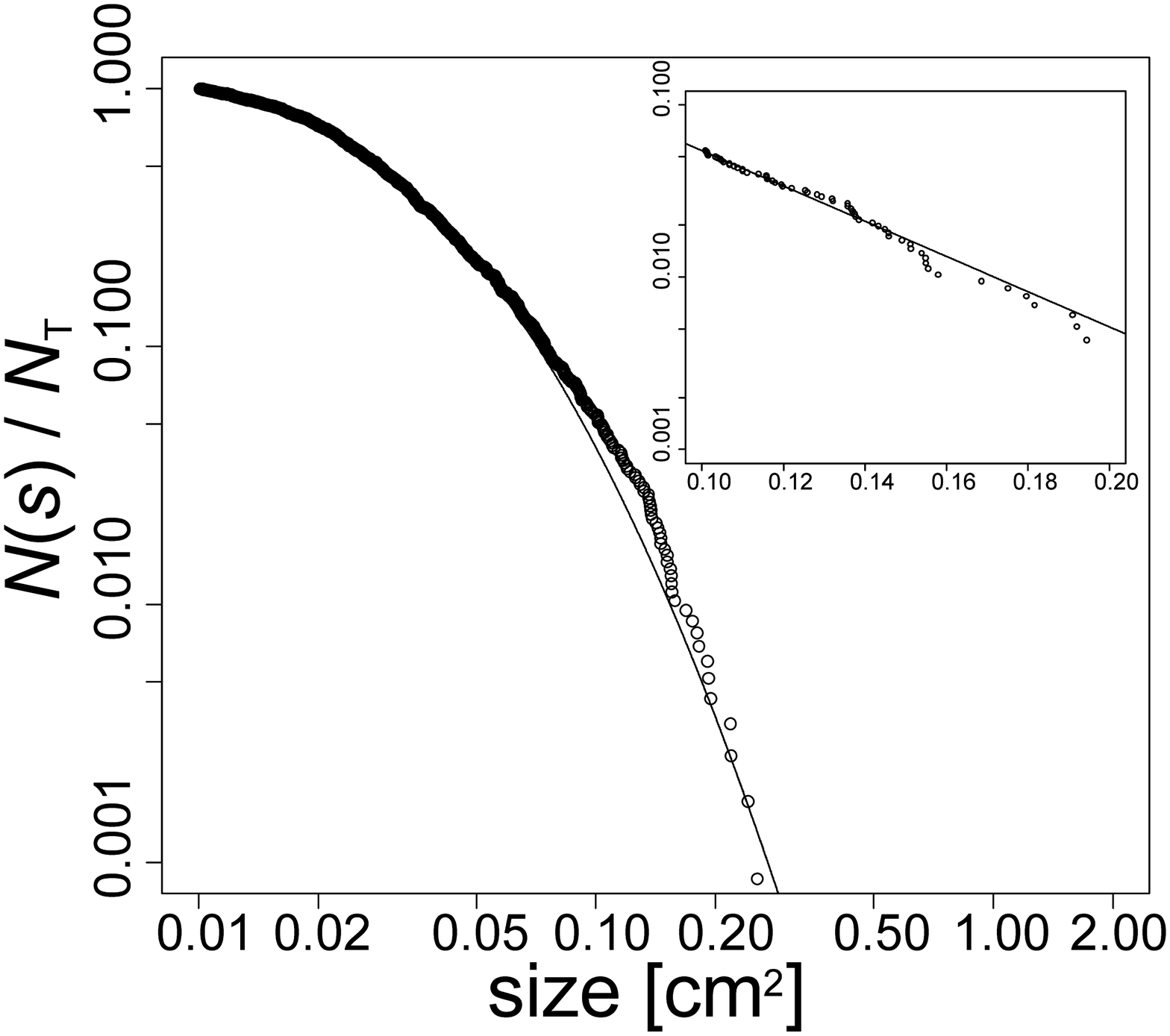}
\label{fig:Fig4a}}
\hfill
\subfigure[]{
\includegraphics*[width=0.42\linewidth]{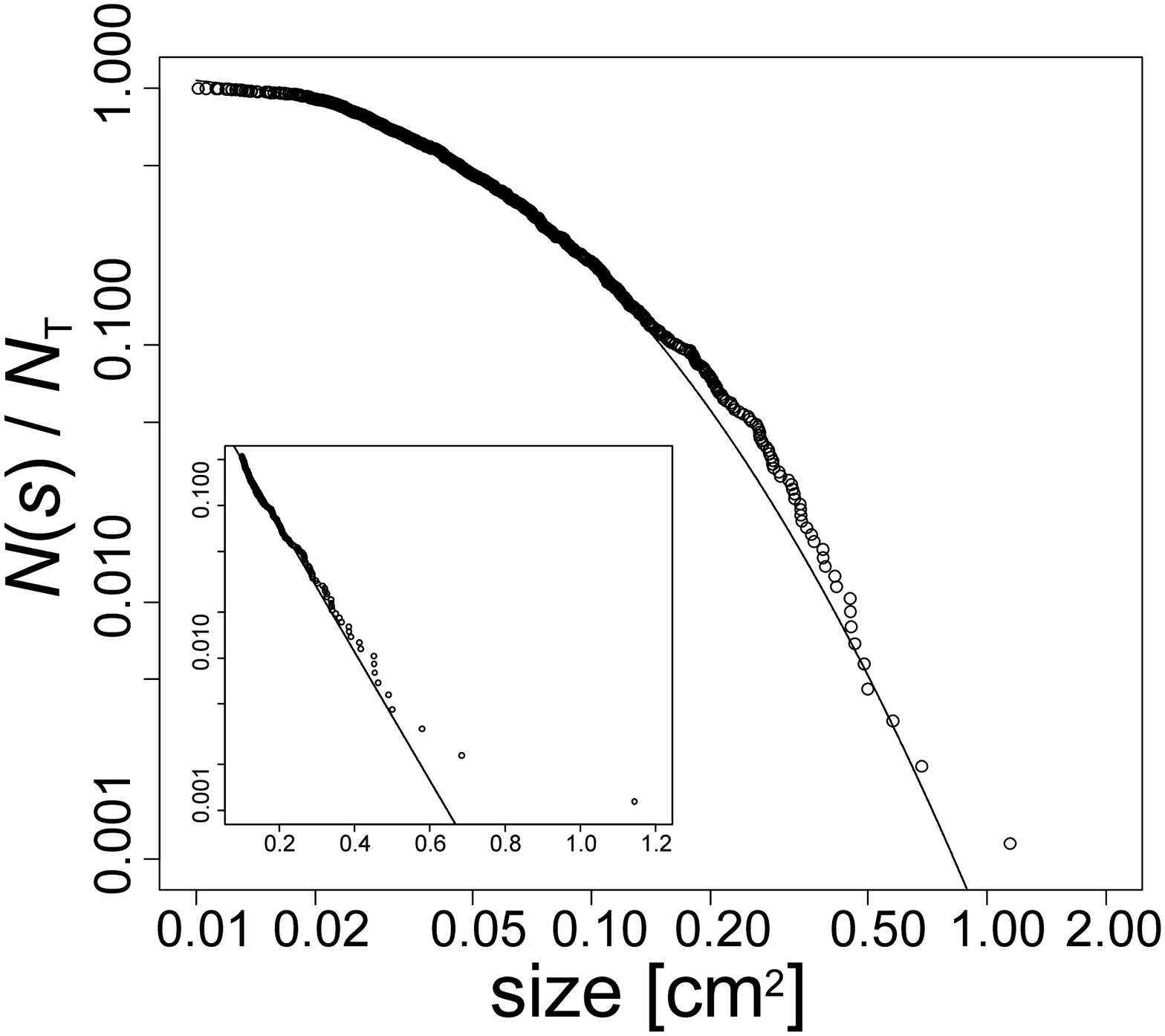}
\label{fig:Fig4b}}
\caption{
Same as Fig. \ref{fig:Fig1} except for the number of chewing
 strokes, i.e., (a) 70 and (b) 35.
The fitting parameters are (a) $N_{\rm{T}} = 1162,
 \bar{s} = 0.0260, \sigma = 0.750$ and $N / N_{\rm{T}} = 1.12$, and
 (b) $N_{\rm{T}} = 873, \bar{s} = 0.0402, \sigma = 0.964$ and
 $N / N_{\rm{T}} = 1.16$.
The fitting parameters for the insets are (a) $A = 0.568$, $B = 0.0425$
 and (b) $A = 0.531$, $B = 0.103$.
}
\label{fig:Fig4}
\end{figure}
\begin{figure}[h]
\centering
\subfigure[]{
\includegraphics*[width=0.42\linewidth]{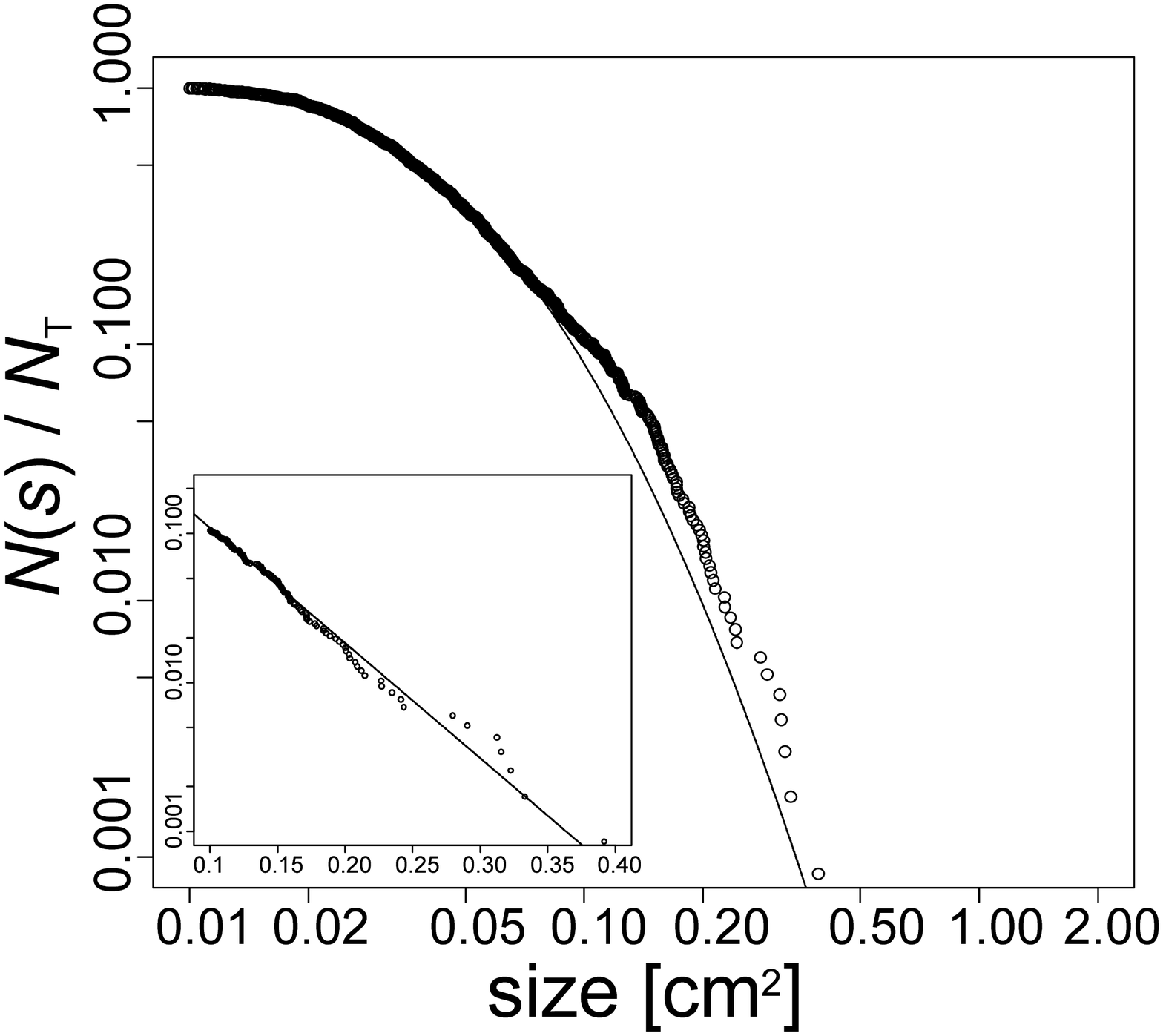}
\label{fig:Fig5a}}
\hfill
\subfigure[]{
\includegraphics*[width=0.42\linewidth]{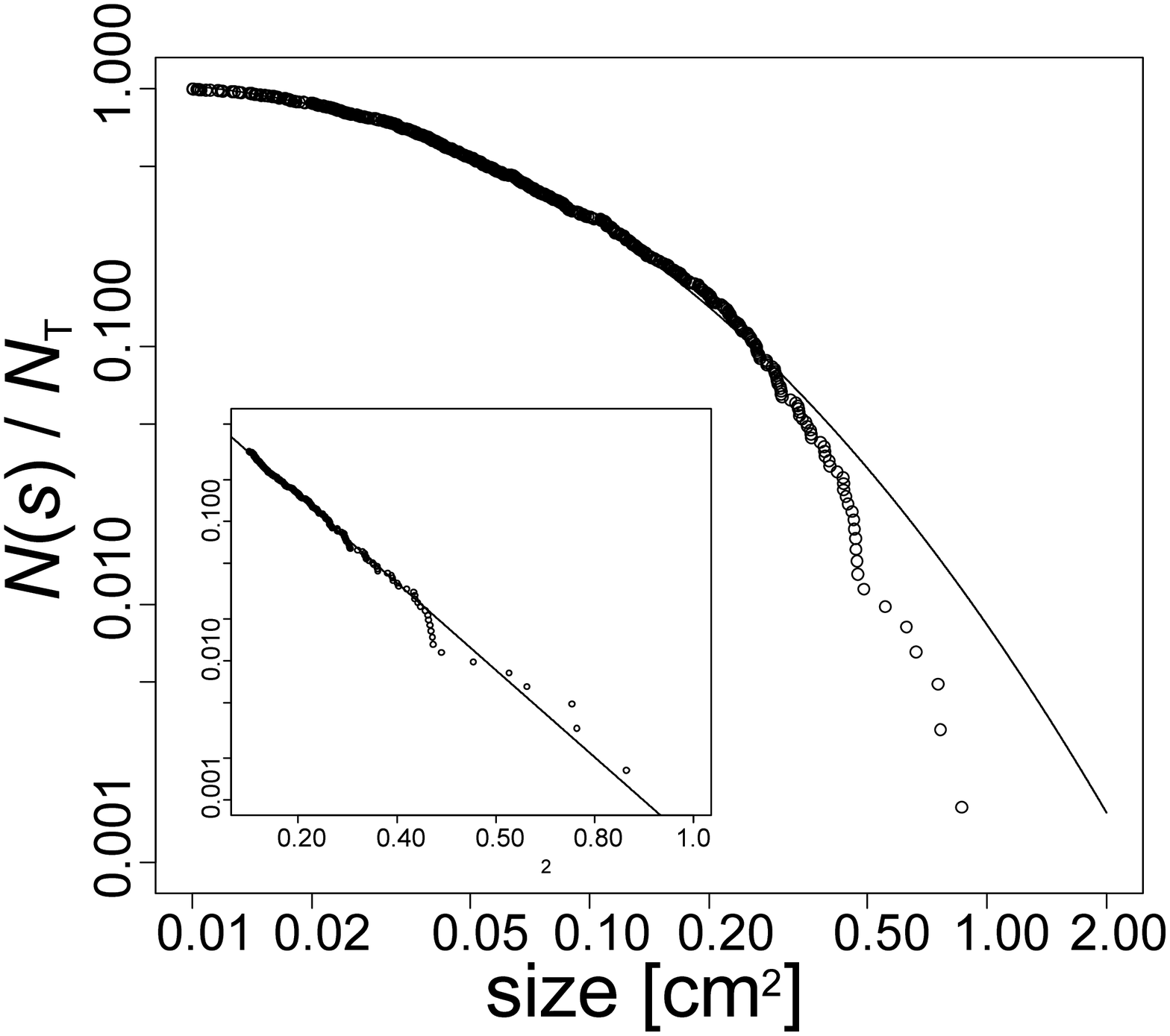}
\label{fig:Fig5b}}
\caption{
Same as Fig. \ref{fig:Fig1} except for the number of chewing
 strokes, i.e., (a) 58 and (b) 29.
The fitting parameters are (a) $N_{\rm{T}} = 1167,
 \bar{s} = 0.0355, \sigma = 0.729$ and $N / N_{\rm{T}} = 1.08$, and
 (b) $N_{\rm{T}} = 611, \bar{s} = 0.0471, \sigma = 1.25$ and
 $N / N_{\rm{T}} = 1.15$.
The fitting parameters for the insets are (a) $A = 0.648$, $B = 0.0562$
 and (b) $A = 0.643$, $B = 0.139$.
}
\label{fig:Fig5}
\end{figure}
\begin{figure}[h]
\centering
\subfigure[]{
\includegraphics*[width=0.42\linewidth]{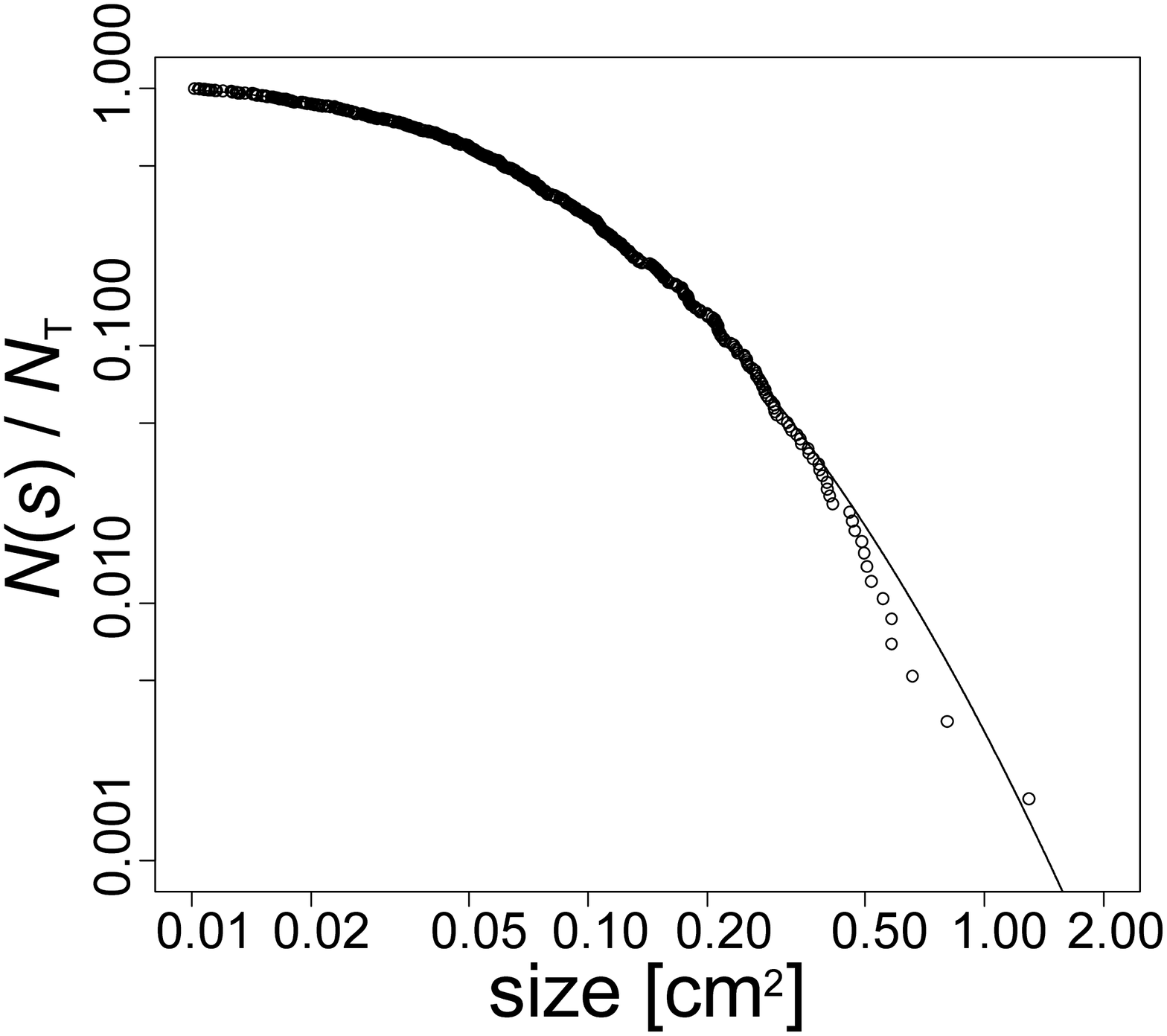}
\label{fig:Fig6a}}
\hfill
\subfigure[]{
\includegraphics*[width=0.42\linewidth]{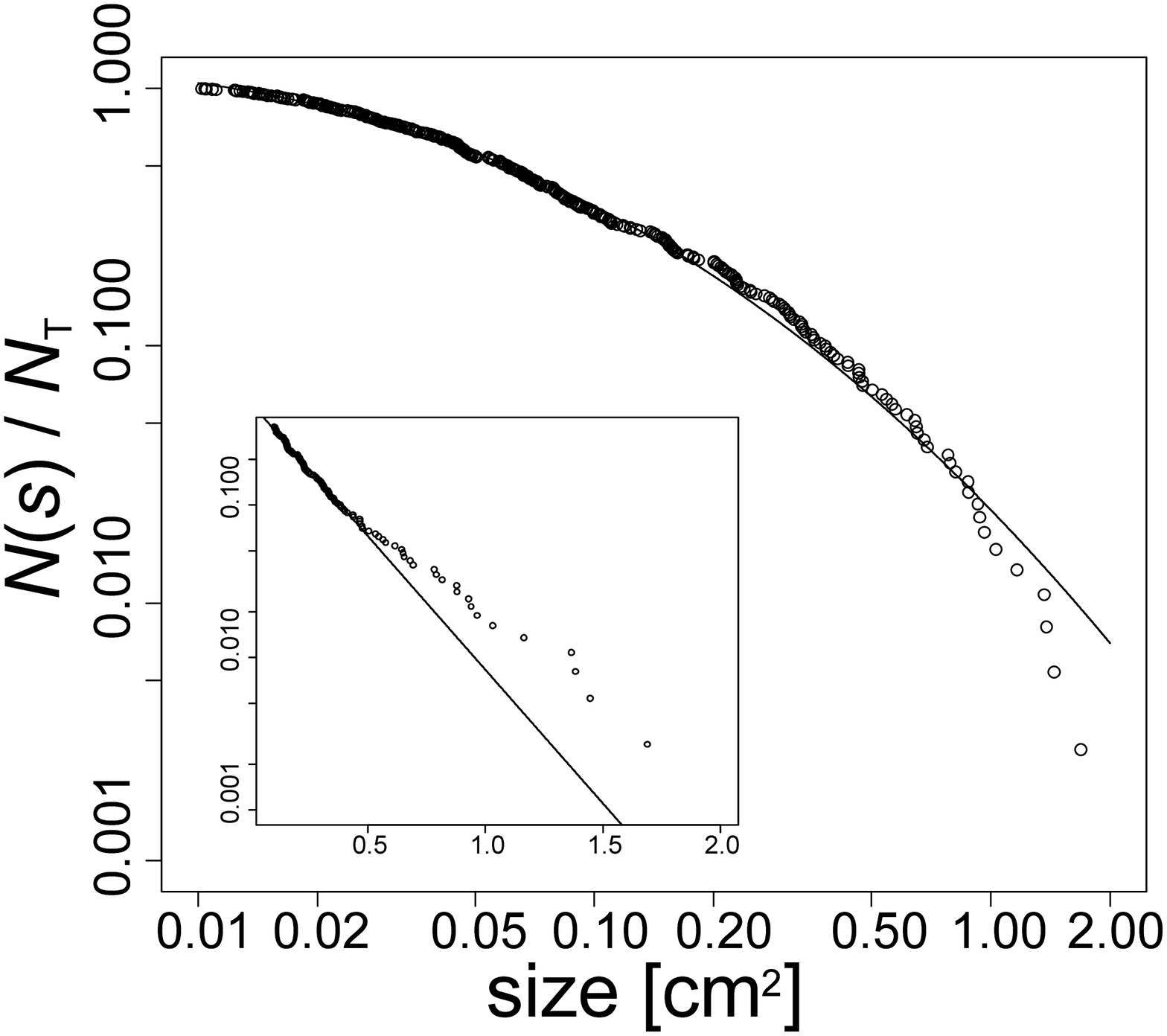}
\label{fig:Fig6b}}
\caption{
Same as Fig. \ref{fig:Fig1} except for the number of chewing
 strokes, i.e., (a) 30 and (b) 15.
The fitting parameters are (a) $N_{\rm{T}} = 576,
 \bar{s} = 0.0595, \sigma = 1.03$ and $N / N_{\rm{T}} = 1.03$, and
 (b) $N_{\rm{T}} = 371, \bar{s} = 0.0384, \sigma = 1.55$ and
 $N / N_{\rm{T}} = 1.30$.
The fitting parameters for the inset are (b) $A = 0.468$, $B = 0.248$.
}
\label{fig:Fig6}
\end{figure}
\begin{figure}[h]
\centering
\subfigure[]{
\includegraphics*[width=0.42\linewidth]{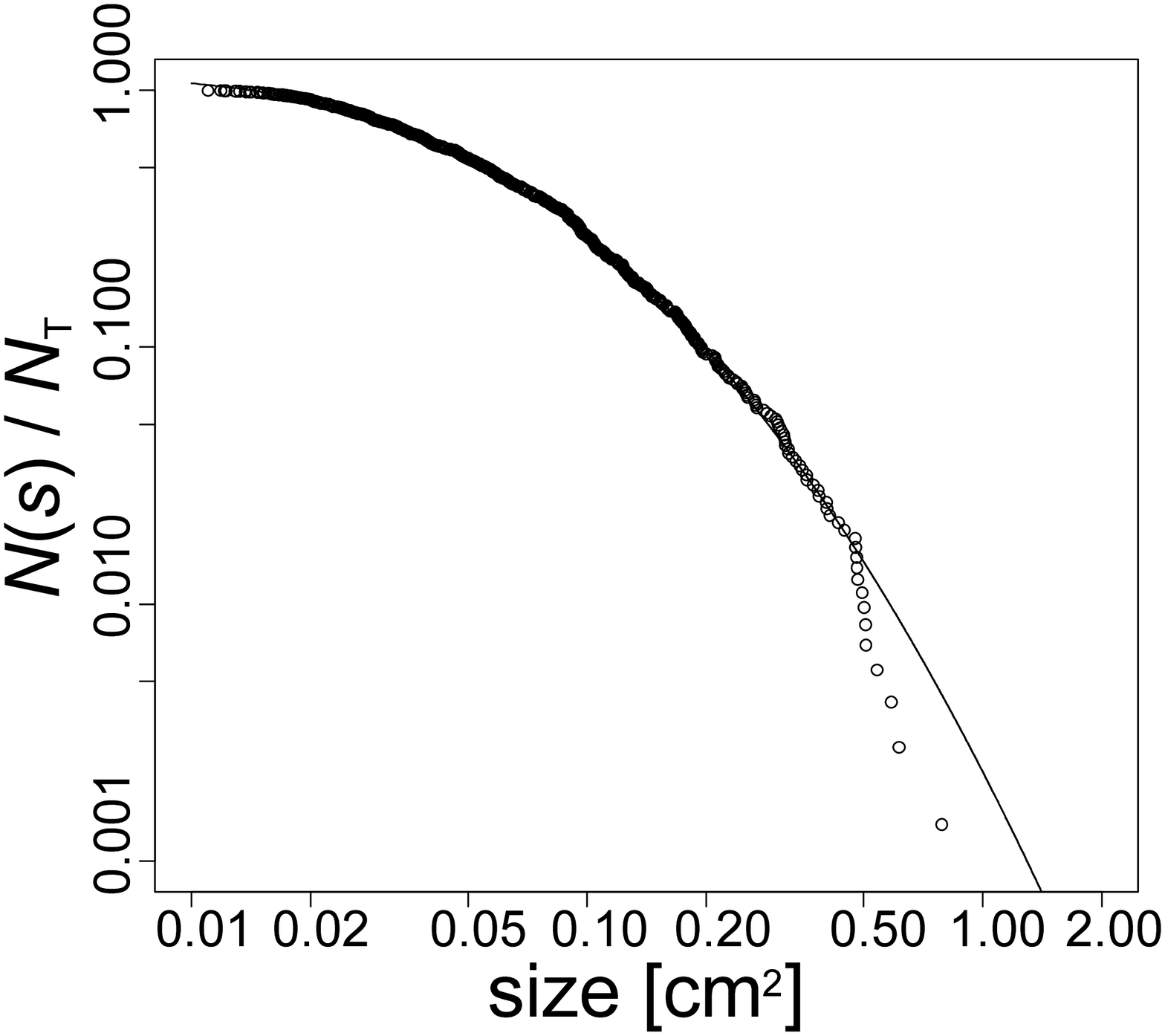}
\label{fig:Fig7a}}
\hfill
\subfigure[]{
\includegraphics*[width=0.42\linewidth]{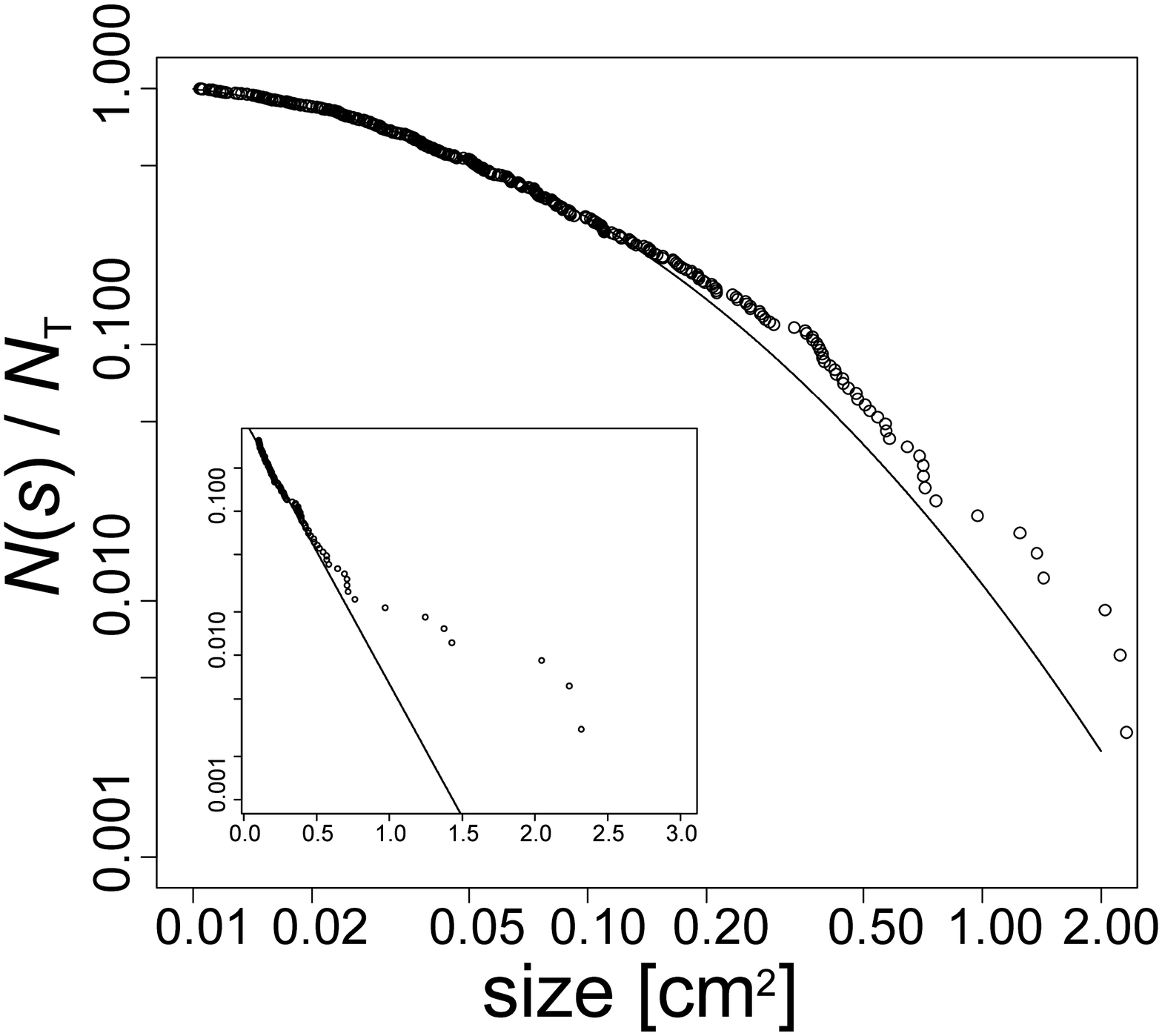}
\label{fig:Fig7b}}
\caption{
Same as Fig. \ref{fig:Fig1} except for the number of chewing
 strokes, i.e., (a) 20 and (b) 10.
The fitting parameters are (a) $N_{\rm{T}} = 722,
 \bar{s} = 0.0468, \sigma = 1.06$ and $N / N_{\rm{T}} = 1.15$, and
 (b) $N_{\rm{T}} = 326, \bar{s} = 0.0444, \sigma = 1.34$ and
 $N / N_{\rm{T}} = 1.34$.
The fitting parameters for the inset are (b) $A = 0.439$, $B = 0.236$.
}
\label{fig:Fig7}
\end{figure}

Figures \ref{fig:Fig1}-\ref{fig:Fig7} show the cumulative
 number of food-fragments.
It should be noted that different figures correspond to different subjects, respectively.
These figures are classified roughly into two groups based on
 fitting curves, namely, single lognormal distribution
 (Figs. \ref{fig:Fig1}, \ref{fig:Fig2}, \ref{fig:Fig6a} and
 \ref{fig:Fig7a}) and lognormal distribution with an exponential
 tail (Figs. \ref{fig:Fig3}, \ref{fig:Fig4}, \ref{fig:Fig5},
 \ref{fig:Fig6b} and \ref{fig:Fig7b}).
For example, as Fig. \ref{fig:Fig1a} shows, the fragment-size
 distribution is nicely approximated by a single lognormal
 distribution given as
\begin{equation}
n(s) = \frac{1}{\sqrt{2 \pi \sigma^2}s} \exp[-\frac{(\log (s/\bar{s})^2)}{2\sigma^2}],
\end{equation}
where $\sigma$ and $\bar{s}$ are the fitting parameters, respectively.
The cumulative form for the lognormal distribution is
\begin{equation}
N(s) = \frac{N_{\mathrm{T}}}{2}(1 - \mathrm{erf}(\frac{\log (s/\bar{s})}{\sqrt{2} \sigma})),
\label{eq:cum_lognormal}
\end{equation}
where $N_{T}$ is the total number of fragments and $\mathrm{erf}(x)$ is the
 error function defined as
 $\mathrm{erf}(x) \equiv (2/\sqrt{\pi})\int_{0}^{x}\exp(-y^2)dy$.
A similar result was reported in our previous study using raw carrot \cite{KKSM06}.

On the other hand, as shown in Figs. \ref{fig:Fig3}, the majority of distributions belong to
 the small region approximated by the lognormal distribution.
However, the tail part of the distributions deviates upwards from the lognormal distribution.
Hence, we propose that the size segregation into small and large food fragment groups
 occured due to physical or other properties of fish sausage.
The insets of Figs. \ref{fig:Fig3} show semi-log plots for the tail part
 (same as Figs. \ref{fig:Fig4}, \ref{fig:Fig5}, \ref{fig:Fig6b}
 and \ref{fig:Fig7b}).
Since the curves on the insets are plotted linearly, the large food fragments were
 fitted to not a lognormal but an exponential distribution,
\begin{equation}
N(s) = A e^{-\frac{s}{B}},
\end{equation}
where $A$ and $B$ are the fitting parameters, respectively.
The physical origin of an exponential distribution of fragmentation is very simple.
If we assume that a fragment-size at each stage of sequential fragmentation is completely
 random, then we obtain the exponential distribution as the fragment-size distribution
 \cite{MS88}.
In the case of our mastication experiments, we think that a similar phenomenon like the above assumption
 happens.

Using the above results, we propose a mechanism for the mastication process of fish sausages
 (see Fig. \ref{fig:Fig8}).
In the first stage of mastication, a unit of fish sausage is broken into a small amount of
 fragments.
Fish sausage is then divided into several more fragments without generating small fragments,
 while raw carrot is divided into many fragments of varied sizes \cite{KKSM06}.
The difference of the fragmentation pattern between fish sausage and raw carrot is derived from
 their physical properties that is fish sausage behaves more plastic than raw carrot in the mouth
If the generated fragments of fish sausage are larger than about $0.1$ cm$^2$, the fragments
 are divided into several fragments again ({\it exponential} in Fig. \ref{fig:Fig8}).
If the generated fragments are smaller, the fragments are ground by the back teeth, and then many,
 heterogeneous and smaller fragments are generated.
After several cycles, the fragments caused by the successive mastication process for fish sausage
will satisfy the two threshold hypothesis suggested by J. B. Hutchings and P. J. Lillford \cite{HL88}.
Here, the chewed food is assembled into a bolus by a complicated movement of the palate and
 the tongue, then just before swallowing \cite{H04}.
According to this proposition, we will conjecture that a fragment-size distribution obtained by
 grinding fish sausage leads to a lognormal distribution ({\it lognormal}  in Fig. \ref{fig:Fig8}).
In fact, Epstein theoretically showed that the fragment-size distribution by grinding solid leads
 to lognormal \cite{Epstein48}.
The proof of this conjecture is one of our future problems.

As mentioned above, we conclude that there are two different kinds of fragment-size distributions,
 i.e., single lognormal distribution and lognormal distribution with exponential tail, because the
 tail part of distribution drifted either downwards or upwards from a lognormal distribution fitted
 to a small fragment group.
This fact indicates that there are differences, such as saliva production and so on, in the mastication
 process for fish sausage among individuals, unlike for raw carrot \cite{KKSM06}.
In fact, individual subjects have shown different behaviour in managing food to terminal swallow \cite{H04}.
The second possibility is that the single lognormal distribution could be close to the lognormal
 distribution with exponential tail in repeated experiments because we must discuss about the
 {\it probability} density distribution.
This is also our second future problem.

\section{Conclusion}

\begin{figure}
\centering
\includegraphics[width=0.7\linewidth]{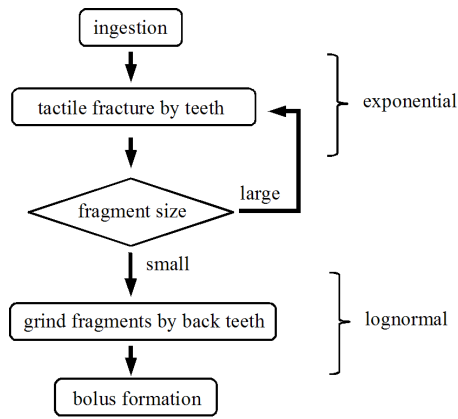}
\caption{A flow chart of the mastication process for fish sausages.}
\label{fig:Fig8}
\end{figure}

In summary, we have studied the fragment-size distribution of masticated viscoelastic food (fish sausage).
We classified the obtained results from 7 humans into two groups, namely, the single lognormal group and
 the lognormal distribution with exponential tail.
The former group shows a similar result as the one in our previous studies using raw carrot.
The latter group is the particular case of masticated fish sausage.
The fragment-size distribution for the latter group shows a double-size-group tendency, i.e.,
 the majority of the distribution belongs to the {\it lognormal} distribution, while the tail behaves
 as the {\it exponential} distribution.
In order to explain this tendency, we suggest a mastication model for fish sausage shown in Fig. \ref{fig:Fig8}.
However, the mastication model remains unfinished, and then we have a lot of further problems, such as
 the size distribution by grinding fragmentation among others.

\begin{acknowledgments}
One of authors (N.K.) is grateful to M. Matsushita, M. Katori, J. Wakita and S. Andraus for many stimulating discussions.
\end{acknowledgments}


\end{document}